\documentclass[reprint,twocolumn,aps,prb,superscriptaddress]{revtex4-2}
\usepackage[T1]{fontenc}
\usepackage[utf8]{inputenc}
\usepackage{lmodern}
\usepackage{graphicx}
\usepackage{amsmath}
\usepackage{xcolor}
\usepackage{float}
\usepackage{hyperref}
\usepackage{braket}
\usepackage{amssymb,amsmath,amsthm, array}
\usepackage{mdframed}
\usepackage{systeme,mathtools}

\def  \bsigma   {\mbox{\boldmath$\sigma $}}
\def  \bnabla   {\mbox{\boldmath$\nabla $}}

\DeclareUnicodeCharacter{0301}{-}

\begin{document}
\renewcommand{\vec}[1]{\mathbf{#1}}
\newcommand{\ii}{\mathrm{i}}
\def\ya#1{{\color{orange}{#1}}}

\title{Random spin-orbit gates in the system of a Topological insulator and a Quantum dot}

\author{S. Wolski}
\affiliation{Department of Physics and Medical Engineering, Rzesz\'ow University of Technology, 35-959 Rzesz\'ow, Poland}
\author{M. Inglot}
\affiliation{Department of Physics and Medical Engineering, Rzesz\'ow University of Technology, 35-959 Rzesz\'ow, Poland}
\author{C. Jasiukiewicz}
\affiliation{Department of Physics and Medical Engineering, Rzesz\'ow University of Technology, 35-959 Rzesz\'ow, Poland}
\author{K. A. Kouzakov}
\affiliation{Faculty of Exact and Natural Sciences, Tbilisi State University, Chavchavadze av.3, 0128 Tbilisi, Georgia}
\author{T. Masłowski}
\affiliation{Department of Physics and Medical Engineering, Rzesz\'ow University of Technology, 35-959 Rzesz\'ow, Poland}
\author{T. Szczepański}
\affiliation{Department of Physics and Medical Engineering, Rzesz\'ow University of Technology, 35-959 Rzesz\'ow, Poland}
\author{S. Stagraczyński}
\affiliation{Institute of Spintronics and Quantum Information, Faculty of Physics, Adam Mickiewicz University, 61-614 Poznań, Poland}
\author{R. Stagraczyński}
\affiliation{Department of Physics and Medical Engineering, Rzesz\'ow University of Technology, 35-959 Rzesz\'ow, Poland}
\author{V. K. Dugaev}
\affiliation{Department of Physics and Medical Engineering, Rzesz\'ow University of Technology, 35-959 Rzesz\'ow, Poland}
\author{L. Chotorlishvili}
\affiliation{Department of Physics and Medical Engineering, Rzesz\'ow University of Technology, 35-959 Rzesz\'ow, Poland}
\affiliation{Faculty of Exact and Natural Sciences, Tbilisi State University, Chavchavadze av.3, 0128 Tbilisi, Georgia}

\date{\today}
\begin{abstract}
The spin-dependent scattering process in a system of topological insulator and quantum dot is studied. The unitary scattering process is viewed as a gate transformation applied to an initial state of two electrons. Due to the randomness imposed through the impurities and alloying-induced effects of band parameters, the formalism of the random unitary gates is implemented. For quantifying entanglement in the system, we explored concurrence and ensemble-averaged Rényi entropy. We found that applied external magnetic field leads to long-range entanglement on the distances much larger than the confinement length. We showed that topological features of itinerant electrons sustain the formation of robust long-distance entanglement, which survives even in the presence of a strong disorder.
\end{abstract}

\maketitle

\section{Introduction}

Quantum computation is an interdisciplinary field with clearly formulated aims and concepts covering fundamental theoretical aspects and practical applications. The feasibility of the experimental realization of quantum technologies depends on  qubits, which are the backbone elements of quantum computation protocols. Formally any two-level quantum system can be viewed as a qubit. There are diverse physical objects considered as prototype qubits. Prospective candidates are semiconductor quantum dots. Since the seminal works  \cite{PhysRevLett.83.4204, PhysRevA.57.120},  semiconductor quantum dots attract the attention of the quantum information community \cite{arakawa2020progress,PhysRevLett.121.110503,PhysRevLett.121.033902,PhysRevB.101.081408,PhysRevB.101.045306,kojima2021probabilistic,PhysRevB.86.085423,PhysRevB.100.174413,PhysRevB.98.104407}. Due to the specific topological properties, quantum dots in graphene,  carbon nanotubes, and topological insulators are less 
vulnerable to decoherence effects \cite{trauzettel2007spin,ezawa2012topological}. On the other hand, in these materials, localized electrons in a quantum dot or itinerant electrons in a topological insulator are influenced by the spin-orbit interaction (SOI), and therefore, SOI may impact the entanglement of the system.

Topological insulators possess fascinating physical features, to mention but a few: coupled spin-charge transport, the spin-momentum locking and specific energy spectrum  \cite{RevModPhys.82.3045,RevModPhys.83.1057,katmis2016high,tokura2019magnetic,PhysRevB.95.094428}.  In the scope of our interest is the specific case of strong bulk topological insulators. We focus on the low-energy excitations characterized by a linear spectrum \cite{PhysRevLett.105.266806}. These features can be important in the  study entanglement between localized electrons in a quantum dot and itinerant surface electrons in a topological insulator. An important problem is the creation of entangled states. A disentangled bipartite system can be entangled through performing particular unitary quantum gate operations. One of the methods for the generation of entanglement is the scattering of two initially disentangled particles. The quantum elastic scattering process can be viewed as a unitary process. Unitary $\hat S$ matrix connects the initial bipartite state with a final state $\ket{\Psi_{AB}^f(\infty)}=\hat S \ket{\Psi_{AB}^i(-\infty)}$. The creation of bipartite entangled pair through scattering was studied in a recent work \cite{PhysRevA.100.022311}. In the present work, we prove that the scattering process entangles the electron localized in the quantum dot with the itinerant electron of the topological insulator. For the sake of realistic discussion, we postulate the effect of the disorder \cite{PhysRevB.96.054440} in the topological insulator, leading to the random spin-orbit interaction constant. Due to the presence of disorder and randomness in the spin-orbit interaction (SOI) constant, the process described below we term as a unitary random spin-orbit gate. 

The experimental feasibility and measuring of entanglement is not a trivial question. Formally to any quantum operator $\mathcal{\hat O}$ one can assign expectation value $Tr(\hat \rho\mathcal{\hat O})$, where $\hat\rho$ is the density matrix of the system. Nonlinear functions of the density matrix i. e., purity $Tr(\hat \rho^2)$ can be measured directly without reconstruction of the whole density matrix 
$\hat \rho$, see \cite{PhysRevA.65.062320,PhysRevLett.89.127902,PhysRevLett.90.167901,PhysRevLett.98.140505} for more details. The direct measurement scheme implies that $n$ identical quantum systems are prepared in the same state $\hat\rho$, and measurements are performed on those multiple copies. Contrary to that, random measurement implies a single copy of $\hat\rho$ only. The concept of the random measurement is based on the random unitary rotation operator $\hat U$ applied to a single copy of the state in question $\hat U\hat \rho_{AB}\hat U^{-1}$, where $\hat\rho_{AB}=\ket{\Psi}\bra{\Psi}_{AB}$ is the density matrix of the bipartite system. The obtained result is averaged over random unitaries, and thus ensemble average of a single random measurement replaces a set of deterministic measurements \cite{PhysRevLett.108.110503,PhysRevLett.120.050406}. For quantum information platforms based on the solid-state systems,  Renyi entropies can be determined from a tomographic reconstruction of the quantum state \cite{haffner2005scalable,PhysRevLett.105.150401,lanyon2017efficient,torlai2018neural}. The second Renyi entropy is measured experimentally \cite{islam2015measuring,kaufman2016quantum}. Therefore interest to the second Renyi entropy is motivated by practical application to SOI systems. Renyi entropies are defined as follows $S^{(n)}(\hat\rho_A)=[1/(n-1)]\log Tr(\hat\rho_A)$, where $\hat\rho_A$ is the reduced matrix $\hat\rho_A=Tr_B(\hat\rho_{AB})$ of the part $A$. Our scope of interest are outcome probabilities
$P(s)=Tr[\hat U\hat\rho_A\hat U^\dagger\mathcal{P}_s]$, where $\mathcal{P}_s=\ket{s}\bra{s}$ are projectors and $\hat U$ is the random unitary matrix. Then Renyi entropy is extracted through the statistical moments
\begin{figure}[h]
\includegraphics[width=1\linewidth]{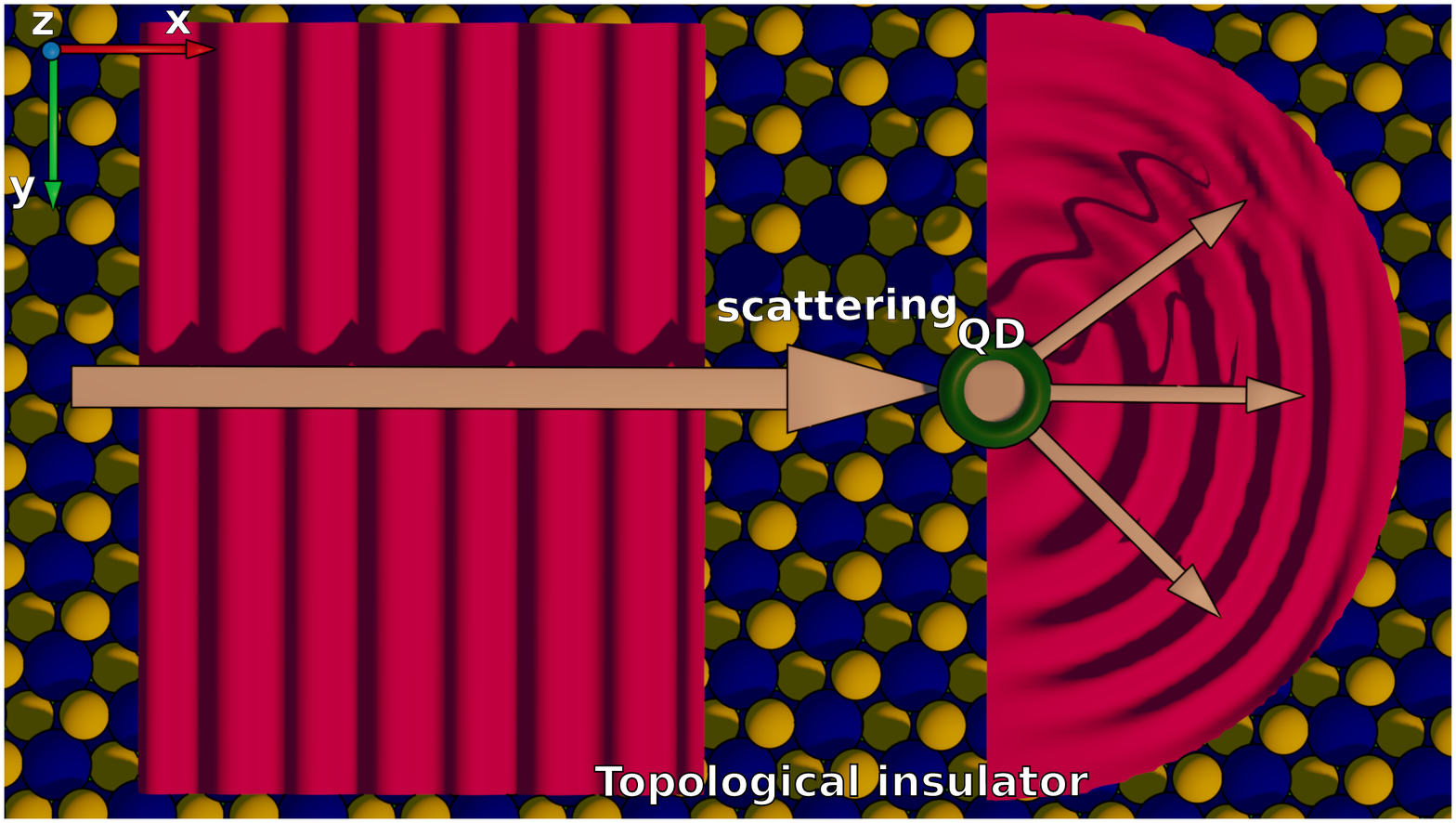}
\caption{Schematics of the system and the scattering process. The itinerant electron from the topological insulator is scattered on the electron localized in the quantum dot (QD). The scattering process depends on the spin of both electrons. Arrows show the direction of motion of the itinerant electron. The momentum of the itinerant electron $k_x=k\cos\theta$, $k_y=-k\sin\theta$. Therefore $\theta=\pi$ and $\theta=0$ correspond to the back and front scattering, while $\theta=\pi/4,\,\theta=-\pi/4$ and $\theta=3\pi/4,\,\theta=-3\pi/4$ correspond to the front and backscattering on the particular angle. All these cases will be analyzed afterward.}\label{pictorial plot}
\label{figure_1}
\end{figure}
\begin{eqnarray}\label{statistical moments}
\langle P(s)^n\rangle=\langle Tr[(\hat U\hat\rho_A\hat U^\dagger)^{\otimes n}\mathcal{P}_s^{\otimes n}]\rangle,
\end{eqnarray} 
where $\langle ...\rangle$ means averaging over disorder. 
Typically random unitaries are implemented using the atomic, molecular, and optical (AMO) toolbox \cite{PhysRevLett.120.050406}. In the present work, we study a system of topological insulator and a quantum dot system and analyze the spin-dependent scattering of the itinerant electron on the electron localized in a quantum dot (see Fig.\ref{pictorial plot}). Thus random unitaries in our proposal are replaced by random SOI elastic scattering process. Let $\ket{\Psi_{AB}^i(-\infty)}$ be the initial wave function of the bipartite system; where $A$ refers the incoming electron (from the topological insulator) and $B$ target electron (from the quantum dot). Consequently $\hat\rho_{AB}(-\infty)=\ket{\Psi_{AB}^i(-\infty)}\bra{\Psi_{AB}^i(-\infty)}$, the initial density matrix. The final density matrix $\hat\rho_{AB}(\infty)=\hat S\hat\rho_{AB}(-\infty)\hat S^{-1}$ is constructed from the wave function of the system after scattering. The scattering matrix $\hat S $ in case of an elastic scattering is viewed as a unitary operation. We analyze the anisotropic and spatially non-uniform distribution of entanglement depending on the scattering process. 

The paper is organized as follows: In section \textbf{II} we describe the model, in section \textbf{III} we introduce the random unitary scattering formalism,
in section \textbf{IV} we present results for concurrence obtained for the deterministic SO coupling constant. The Rényi entropy is discussed in section \textbf{V}. Section \textbf{VI} concludes the work. Technical details of derivations are presented in the appendix.

\section{Model}
\label{sec:Model}

The topological surface states differ from the slab band structures and are termed as the "Dirac cone".
The surface-state bands are doubly degenerate. Two surface states are characterized by the
opposite spin helicities. For more details about spectral properties of topological insulators we refer to the work \cite{PhysRevLett.105.266806}.
We consider the system with a single quantum dot on top of a topological insulator. The Hamiltonian of this system has the following form:
\begin{eqnarray}
\label{The total Hamiltonian of the system}
&& \hat H_{tot}=\hat H_{I}+\hat H_{D}+\hat V.
\end{eqnarray}
Here $\hat H_{I}=-iv\, \hat{\bsigma }_A\cdot \bnabla $ is the 2D Hamiltonian of electrons at the surface of the topological insulator and $\hat{\bsigma }_A$ is the spin matrix acting on their spins. In what follows, we assume that the velocity parameter $v$ is a random variable with a mean value $\langle v\rangle$. The main source of randomness is the disorder due to impurities leading to random SO interaction constant \cite{mel1972influence, PhysRevB.67.161303} in the bulk of TI. The alloying-induced effect of band parameter variation in such topological insulators like Bi$_{1-x}$Sb$_x$ or Bi$_{2-x}$Sb$_x$Te$_{3-y}$Se$_y$ \cite{ando2013topological} also contributes to randomness. 

 The interaction between itinerant electrons is neglected in the single electron description of the topological insulators. However, many physical properties can be described by this simplified model. Concerning electronic and spin correlations, we refer to the works \cite{PhysRevB.93.235112,PhysRevB.95.205120,PhysRevB.98.045133,PhysRevB.96.041104,PhysRevLett.120.096601,PhysRevB.100.235419}. The impact of interaction between itinerant electrons is material specific \cite{PhysRevB.88.045206,PhysRevB.93.205442} and is significant at relatively high energies $E>0.1$ eV
\cite{shen2012topological}. In what follows, our description is limited only to the low-energy excitations $E<0.1$eV.  Therefore, the interaction effect between itinerant electrons is beyond the scope of the present work.

For certain realization of the parameter $v$, the eigenstates and eigenenergies of the surface electrons in the system without quantum dot read:
\begin{eqnarray}\label{eigen states of the topological insulator}
&&\psi_{T,\sigma}(\textbf{r})
=\frac{e^{i\textbf{k}\cdot \textbf{r}}}{\sqrt{2}} \begin{pmatrix} 1 \\ \pm k_+/k \end{pmatrix},\hskip0.3cm
E =\pm vk,
\end{eqnarray}
where $k_+=k_x+ik_y$, $k=\sqrt{k_x^2+k_y^2}$ and $\textbf{r}=(x,y)$.

The Hamiltonian of electron localized in the quantum dot (QD) has the form:
\begin{eqnarray}\label{Hamiltonian of the electron localized}
&&\hat H_{D}=-B\hat\sigma_B^z-\frac{\hbar^2}{2m}\vec\nabla^2+\frac{1}{2}m\omega_0^2r^2,
\end{eqnarray}
where $\omega_0$ is the frequency of electron oscillation in QD.  
The external magnetic field applied locally to the quantum dot allows to freeze (strong field) or relax (weak field)
the spin of the localized electron $\hat\sigma_B^z$ depending on the value of Zeeman energy $B\equiv \hbar\gamma_eB$.
The lowest eigenstate of the localized electron has the form 
\begin{eqnarray}\label{confined in the quantum dot}
&&\psi_{D}(\textbf{r})=\frac{1}{l_B\sqrt{\pi}}\, \exp\left(-\frac{x^2+y^2}{2l_B^2}\right) ,
\nonumber\\
&& l_B^2=l_0^2/\sqrt{1+B^2e^2l_0^4/4\hbar^2}.
\end{eqnarray}
Here $l_0=(\hbar/m\omega_0)^{1/2}$ is the confinement length \cite{PhysRevB.82.045311}.
 The magnetic field can be applied locally to the quantum dot, e.g., through spin-polarized scanning tunneling microscopy (SP-STM) \cite{PhysRevResearch.3.043185,willke2019tuning}.  Therefore the magnetic field does not affect the electronic structure of the topological insulator.

The last term in Eq.(\ref{The total Hamiltonian of the system}) describes the interaction between localized and surface electrons and has the form \cite{PhysRevB.89.075426}:
\begin{eqnarray}\label{interaction between localized and itinerant electrons}
&&\hat V =J\, \hat\bsigma _A\cdot \hat\bsigma _B\; \delta\left(\textbf{r}_1-\textbf{r}_2\right).
\end{eqnarray}
The coupling constant $J$ is determined by $J=4RT^2/U$, 
whereas $U$ and $T$ are the Coulomb interaction and electron hopping between topological insulator and QD:
\begin{eqnarray}\label{localized and itinerant}
&&U=\frac{e^2}{4\pi\varepsilon_0}\int\frac{\vert \psi_{T,\sigma}(\textbf{r}_1)\vert^2\, 
\vert \psi_{D,\sigma}(\textbf{r}_2)\vert^2}
{\vert\textbf{r}_1-\textbf{r}_2\vert+\delta}\, d\textbf{r}_1\, d\textbf{r}_2,
\nonumber \\
&& T=E\int\psi_{T,\sigma}^\dag (\textbf{r})\, \psi_{D,\sigma}(\textbf{r})\, d\textbf{r},\,\,\,
J=\frac{4RT^2}{U}.
\end{eqnarray}
Here $\hat\bsigma _B$ refers to the spin operator of the quantum dot and $\delta $ is the cut-off of electron-electron interaction, $R$ is the radius of the quantum dot.
  
As we see from Eq.(\ref{interaction between localized and itinerant electrons}) and Eq.(\ref{localized and itinerant}) the interaction constant $J$ depends on several parameters, such as velocity and wave vector of the itinerant electrons, as well as the QD size. Through the coupling parameters one can tune the strength of interaction. In what follows, we consider mean value of the constant $J$ by replacing $E\equiv\langle E\rangle=\langle vk\rangle$ in Eq.(\ref{localized and itinerant}) and  set the dimensionless parameters $\hat H_{tot}\equiv\hat H_{tot}/J$. \\

\section{Locally randomized gates}
\label{sec:gates}

 In experiments with entangled states in topological insulators aiming at the use of quantum gates, any summation over multiple states should be eliminated. This can be done by collimating electron flux, like in optical experiments with light. In this case, we can get a single incoming electron state with a specific momentum value ${\bf k}$, for which one can use our calculation results. According to the Landau theory of Fermi liquid, the states at the Fermi level with $k\sim k_F$ mainly contribute to the transport properties. A preferable direction of electron motion depends on the direction of an applied electric field. Therefore momentum of electrons contributing to the effect is well defined by the value and direction, and integration over the whole Fermi surface is unnecessary.

Before analyzing  Rényi entropy, we describe random unitary gates. Despite some similarity to the Kondo scattering problem, our model differs significantly from the Kondo problem, i. e., from the spin-flip electron scattering on a localized magnetic moment   \cite{tsvelick1983exact, RevModPhys.55.331}. 
Kondo Hamiltonian $\hat H_{sd}$ can be derived from Anderson's impurity model after considering the following assumptions \cite{altland2010condensed}: Hubbard $U$ and the energy of the impurity $E_d$ should be significantly in excess as compared to the energy of itinerant electron $E$. This assumption oversimplifies the Kondo problem because the scattering term can be absorbed into a shift of the single-particle energy of the itinerant band. Besides, the large Hubbard repulsion energy $U$ implies a short distance between itinerant and localized electrons. The short distance between electrons is irrelevant if we look for robust long-distance entanglement in the system. In the present work, we propose the formulation of the scattering problem in a more general form in the spirit of the Lippmann–Schwinger integral equation. As opposed to the Kondo problem, the interaction between itinerant and localized electrons, in our case, depends on the energy of the itinerant electron (momentum of the Dirac massless particle $\textbf{k}$). What is even more critical,  the entanglement also depends on $\textbf{k}$. Therefore the effect of our interest cannot be captured within the framework of the Kondo problem.

Two types of processes are in the scope of interest [i] the scattering when both spins $\hat\sigma_A$, $\hat\sigma_B$ are flipped, [ii] only spin of itinerant electron $\hat\sigma_A$ is flipped (spin of the localized electron $\hat\sigma_B$ can be frozen by applying strong magnetic field). The initial wave function of the bipartite system is a product of two wave functions $\psi_{T,\sigma}(\textbf{r}_A)=\psi(\textbf{r}_A)\left(\alpha\vert 0\rangle_A+\beta\vert 1\rangle_A\right)$ and $\psi_{D,\sigma}(\textbf{r}_B)=\psi(\textbf{r}_B)\vert 0\rangle_B$. In what follows, for brevity, we use the notations $\textbf{r}_A\equiv\textbf{r}$ and  $\textbf{r}_B\equiv\textbf{r}'$.
Let us assume that the magnetic field  is not zero and localized electron after scattering stays in the ground state, while spin is flipped. Under this constraint, the scattering process involves two states of localized electron $\psi_{D,0}(\textbf{r})=\psi_D(\textbf{r})\vert 0\rangle$ (spin-up, $|0\rangle\equiv\ket{\uparrow}$) and  $\psi_{D,1}(\textbf{r})=\psi_D(\textbf{r})\vert 1\rangle$ (spin-down, $|1\rangle\equiv\ket{\downarrow}$) with the respective energies $\varepsilon_0$ and $\varepsilon_1=\varepsilon_0+2B$ (hereafter, we set $\varepsilon_0=0$). Neglecting the exchange effects, the wave function of the two-electron system can be presented in the following general form:
\begin{equation}\label{two-electron system}
\Psi_{\sigma_1\sigma_2}(\textbf{r}_1,\textbf{r}_2)=\psi_{T,0}^{(+)}(\textbf{r}_1)\psi_{D,0}(\textbf{r}_2)+\psi_{T,1}^{(+)}(\textbf{r}_1)\psi_{D,1}(\textbf{r}_2).
\end{equation}
Here $\psi_{T,0}^{(+)}$ and $\psi_{T,1}^{(+)}$ are the two-component spinors
\begin{equation}\label{two-component spinors}
\psi_{T,0}^{(+)}(\textbf{r})= \begin{pmatrix} \phi_0(\textbf{r}) \\  \chi_0(\textbf{r}) \end{pmatrix}, \qquad 
\psi_{T,1}^{(+)}(\textbf{r})= \begin{pmatrix} \phi_1(\textbf{r}) \\  \chi_1(\textbf{r})
\end{pmatrix}.
\end{equation}
Spinors in the above equation are the solution of the following system of coupled integral equations:
\begin{widetext}
\begin{align}
\begin{pmatrix} \phi_0(\textbf{r}) \\ \chi_0(\textbf{r}) \end{pmatrix} &=\frac{e^{i\textbf{kr}}}{\sqrt{2}}
\begin{pmatrix} 1 \\  k_+/k \end{pmatrix} +\int d^2r'\,\hat{G}^{(+)}(\textbf{r},\textbf{r}';E)
\hat{V}_{00}(\textbf{r}')\begin{pmatrix} \phi_0(\textbf{r}') \\  \chi_0(\textbf{r}') \end{pmatrix}
+\int d^2r'\,\hat{G}^{(+)}(\textbf{r},\textbf{r}';E)
\hat{V}_{01}(\textbf{r}')\begin{pmatrix} \phi_1(\textbf{r}') \\  \chi_1(\textbf{r}') \end{pmatrix},\nonumber
\end{align}
\begin{align}\label{coupled integral equations second}
\begin{pmatrix} \phi_1(\textbf{r}) \\ \chi_1(\textbf{r}) \end{pmatrix} &=\int d^2r'\,\hat{G}^{(+)}(\textbf{r},\textbf{r}';E-2B)\hat{V}_{10}(\textbf{r}')\begin{pmatrix} \phi_0(\textbf{r}') \\  \chi_0(\textbf{r}') \end{pmatrix}+\int d^2r'\,\hat{G}^{(+)}(\textbf{r},\textbf{r}';E-2B)
\hat{V}_{11}(\textbf{r}')\begin{pmatrix} \phi_1(\textbf{r}') \\  \chi_1(\textbf{r}') \end{pmatrix},
\end{align}
\end{widetext}
where $E=vk$ is the energy of itinerant electron and the Green's function is given by
\begin{widetext}
\begin{equation}\label{The Green function}
\hat{G}^{(+)}(\textbf{r},\textbf{r}';E)=\langle\textbf{r}|\frac{1}{E-\hat{H}_I+i0}|\textbf{r}'\rangle=-\frac{E}{2\pi v^2}\left[K_0(-iE|\textbf{r}-\textbf{r}'|/v)\hat{I}+
K_1(-iE|\textbf{r}-\textbf{r}'|/v)\hat{\sigma}^x\right],
\end{equation}
\end{widetext}
where $K_{0,1}$ are the modified Bessel functions (the Macdonald functions). The explicit form of the matrix elements $\hat V_{nm}$ are presented in the appendix.
We solve Eq.(\ref{coupled integral equations second})
in the Born approximation and after cumbersome calculations obtain expression of bipartite wave function after scattering:
\begin{widetext}
\begin{eqnarray}\label{after cumbersome calculations}
&&\ket{\Psi_{\sigma_1\sigma_2}(\mathbf{r},\mathbf{r}')}=\frac{1}{C_0}\left\lbrace 
C_1\psi_D(\rho')\ket{0}_A\ket{0}_B+C_2\psi_D(\rho')\ket{0}_A\ket{1}_B+C_3\ket{1}_A\ket{0}_B+
C_4\psi_D(\rho')\ket{1}_A\ket{1}_B\right\rbrace. 
\end{eqnarray}
\end{widetext}
Here we introduced the following notations:
\begin{widetext}
\begin{eqnarray}\label{here one}
&& C_1\left(\rho,\rho',\varphi\right)=\frac{\psi_D(\rho')}{2\pi\sqrt{2}}e^{ik\rho\cos(\varphi)}\left(2\pi+k_Jk\left(\tilde{A}_{00}(\rho,\varphi)+e^{i\theta}\tilde{A}_{10}(\rho,\varphi)\right)\right),\nonumber\\
&&C_2\left(\rho,\rho',\varphi\right)=-\frac{\psi_D(\rho')}{2\pi\sqrt{2}}e^{ik\rho\cos(\varphi)}\left(2k_J(k-2k_B)\tilde{A}_{0B}(\rho,\varphi)\right)e^{i\theta},\nonumber\\
&&C_3\left(\rho,\rho',\varphi\right)=\frac{\psi_D(\rho')}{2\pi\sqrt{2}}e^{ik\rho\cos(\varphi)}\left(2\pi e^{i\theta}+k_Jk\left(e^{i\theta}\tilde{A}_{00}(\rho,\varphi)+\tilde{A}_{10}(\rho,\varphi)\right)\right),\nonumber\\
&&C_4\left(\rho,\rho',\varphi\right)=-\frac{\psi_D(\rho')}{2\pi\sqrt{2}}e^{ik\rho\cos(\varphi)}\left(2k_J(k-2k_B)\tilde{A}_{1B}(\rho,\varphi)\right)e^{i\theta},\nonumber\\
&&C_0=\sqrt{\vert C_1\vert^2+\vert C_2\vert^2+\vert C_3\vert^2+\vert C_4\vert^2},\,\,\,
\tilde{A}_{0B}(\rho,\varphi)=A_{lB}(\rho,\varphi)e^{-ik\rho\cos(\varphi)},
\end{eqnarray}
\end{widetext}
where $\rho$ and $\varphi$ are polar coordinates. Dimensionless parameters are introduced through the following notations $k_+/k=e^{-i\theta}$, $J/v=k_J$, $E\equiv E/J$, $E_B=E-2B$, $E/v=k$, $E_B/v=k-2k_B$, $k_B\equiv B/v$, and $l=0,1$.
In the coefficients $A_{lB}(\mathbf{r})$, indexes $l=0,1$ define zero and the first-order modified Bessel functions and $B=0,1$ indicates on the zero and nonzero magnetic field problems. Explicit forms of coefficients are involved and are presented in the appendix. Eq.(\ref{after cumbersome calculations}) contains random parameter $v$ over which we do ensemble average.

\section{Concurrence}
\label{sec:Concurrence}

Taking into account the initial density matrix 
\begin{eqnarray}\label{the initial density matrix}
\hat\rho_0=\left(\psi_{T,\sigma}(\textbf{r})\psi_{D}\right)
\left(\psi_{T,\sigma}(\textbf{r})\psi_{D}\right)^\dagger\otimes\ket{0}\bra{0}_B,
\end{eqnarray}
and the wave function of the system after scattering Eq.(\ref{after cumbersome calculations}), we define two random unitary matrices through following formulae 
\begin{eqnarray}\label{through following formulae first}
&&\hat\varrho=\ket{\Psi_{\sigma_1\sigma_2}(\mathbf{r},\mathbf{r}')}\bra{\Psi_{\sigma_1\sigma_2}(\mathbf{r},\mathbf{r}')}=\nonumber\\
&&\frac{1}{Tr\left(\hat U\hat\rho_0\hat U^\dagger\right)}\hat U\hat\rho_0\hat U^\dagger,
\end{eqnarray}
\begin{eqnarray}\label{through following formulae second}
&&\hat\varrho_A=\frac{1}{Tr\left(\hat U_A\hat\rho_0\hat U_A^\dagger\right)}\hat U_A\hat\rho_0\hat U_A^\dagger.
\end{eqnarray}
Here index $A$ means that operator $\hat U_A$ acts only on the qubit $A$ while the spin of the localized electron is frozen by applied strong magnetic field. 
\begin{figure*}[htbp]
\includegraphics[width=0.9\linewidth]{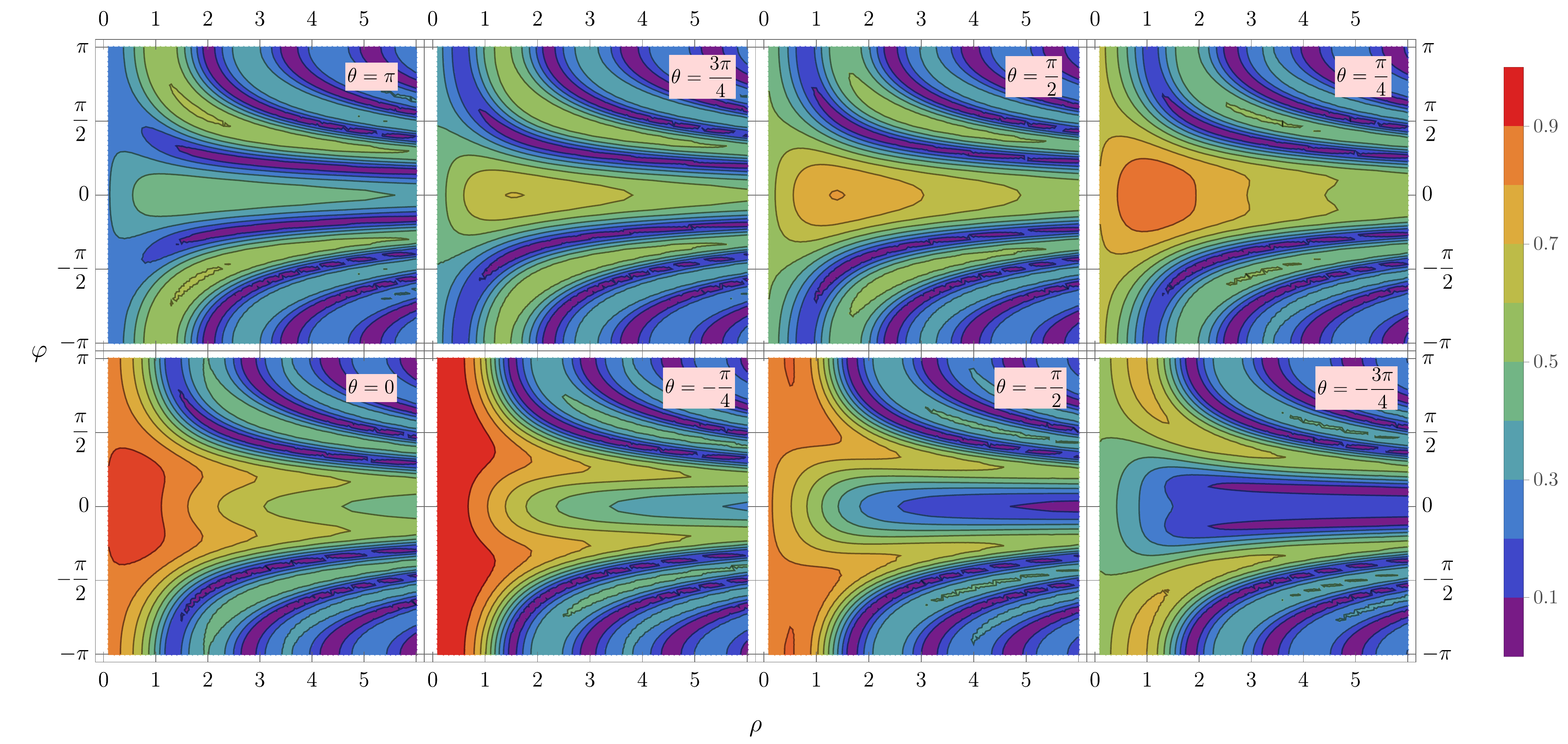}
\caption{The concurrence in the absence of the disorder, plotted for the values of the parameters $E\equiv E/J=1$, $E/v=k=1$, $k_B\equiv B/v=0$.}
\label{figure_2}
\end{figure*}
We exploit definition of concurrence \cite{PhysRevLett.80.2245}
$\mathcal{C}=\vert\bra{\psi}\hat\sigma_y\otimes\hat\sigma_Y\ket{\psi^*}\vert$, where $\hat\sigma_y\otimes\hat\sigma_Y$ is the direct product of Pauli matrices and calculate entanglement between two electrons after scattering:
\begin{widetext}
\begin{eqnarray}\label{entanglement between two electrons}
\mathcal{C}=\frac{2}{C_0^2}\vert C_2^*\left(\rho,\rho',\varphi\right)C_3^*\left(\rho,\rho',\varphi\right)-C_1^*\left(\rho,\rho',\varphi\right)^*C_4^*\left(\rho,\rho',\varphi\right)^*\vert. 
\end{eqnarray}
\end{widetext}
In what follows we explore concurrence in absence and presence of disorder. 

\begin{figure*}[htbp]
\includegraphics*[width=0.75\linewidth]{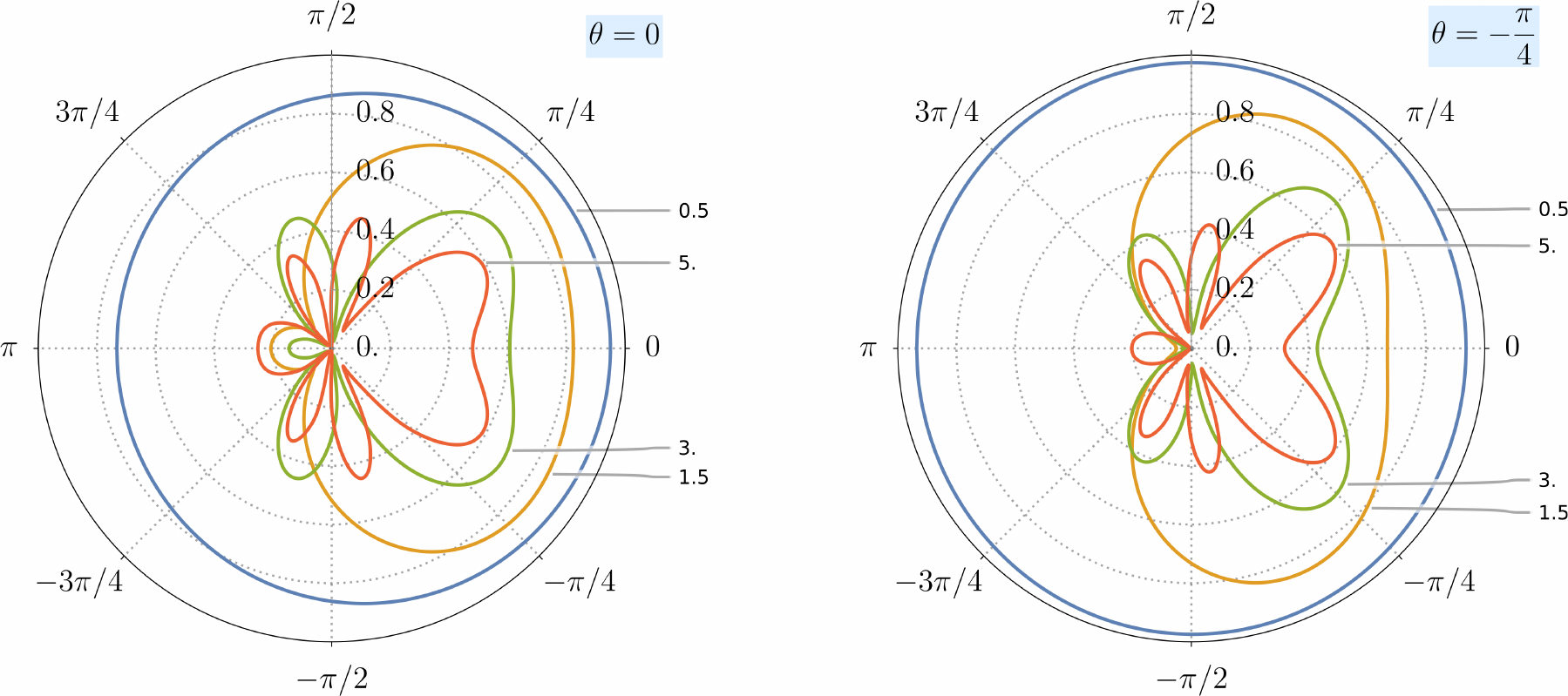}
\caption{The concurrence in the absence of the disorder plotted in the polar coordinate system for the values of the parameters $E\equiv E/J=1$, $E/v=k=1$, $k_B\equiv B/v=0$. Two particular cases of the maximal concurrence are selected: i. e., an electron from a topological insulator after the scattering passing along the $X$ axis ($\theta=0$),  $k_x>0$, $k_y=0$, or scattered back on the angle $\theta=-\pi/4$, $k_x<0$, $k_y>0$. Curves of different colors correspond to the different distances between the scattered electron and the center of the quantum dot. The distance $\rho/l_0$ is measured in the units of confinement length $l_0$. Angle $\varphi$ characterizes anisotropy of the spatially resolved distribution of concurrence.}
\label{figure_3}
\end{figure*}

 As we see from Eq.(\ref{here one}), Eq.(\ref{entanglement between two electrons}), concurrence depends on several factors such as: 
distance of the confined electron $\varrho'$ from the center of the quantum dot (which we set to be equal to the localization length $\varrho'=l_0=1$), the distance between itinerant electron and center of the quantum dot $\varrho$, applied magnetic field $B$ and the direction of the momentum of itinerant electron (characterized by angle $\theta$) after the scattering process $k_+=k_x+ik_y$, $k=\sqrt{k_x^2+k_y^2}$, $e^{-i\theta}=k_+/k$. The value of the concurrence is averaged over the angular distribution of the confined electron (see integration over $\phi'$ in Eq.(29)). On the other hand, concurrence is anisotropic with respect to the angular distribution of the density of scattered itinerant electron (Concurrence depends on the angle $\phi$). For the sake of simplicity, we redefine $\varrho$ in terms of the distance between the border of the quantum dot and scattered itinerant electron $\rho=\varrho-l_0\geqslant 0$, meaning that the electron from the topological insulator cannot penetrate inside the quantum dot. In Fig.\ref{figure_2} is plotted concurrence as a function of the distance between itinerant electron and border of the quantum dot $\rho$. The distance is measured in the units of the confinement length $l_0$. At first, we assume that the magnetic field is not applied $B=0$. As we see, concurrence depends on the direction of the scattered electron ($\theta$) and is anisotropically distributed (dependence on the angle $\varphi$ is shown). Concurrence is maximal for electrons moving along the $X$ axis ($\theta=0$),  $k_x>0$, $k_y=0$, or scattered back on the angle $\theta=-\pi/4$, $k_x<0$, $k_y>0$. We also present these two cases of particular interest in the form of a polar plot in Fig.(\ref{figure_3}). 

As we see in Fig.(\ref{figure_3}), concurrence decays with the distance from the center of the quantum dot (The distance $\rho/l_0$ is measured in the units of confinement length $l_0$). We clearly see anisotropy of the spatially resolved distribution of concurrence. In particular, concurrence depends on the polar angle of the scattered electron $\varphi$. This fact means that not only the direction of the momentum of the scattered electron $\textbf{k}$, but the position of the scattered electron on the $XOY$ plane is an essential factor for the concurrence.

\begin{figure*}[htbp]
\includegraphics*[width=0.9\linewidth]{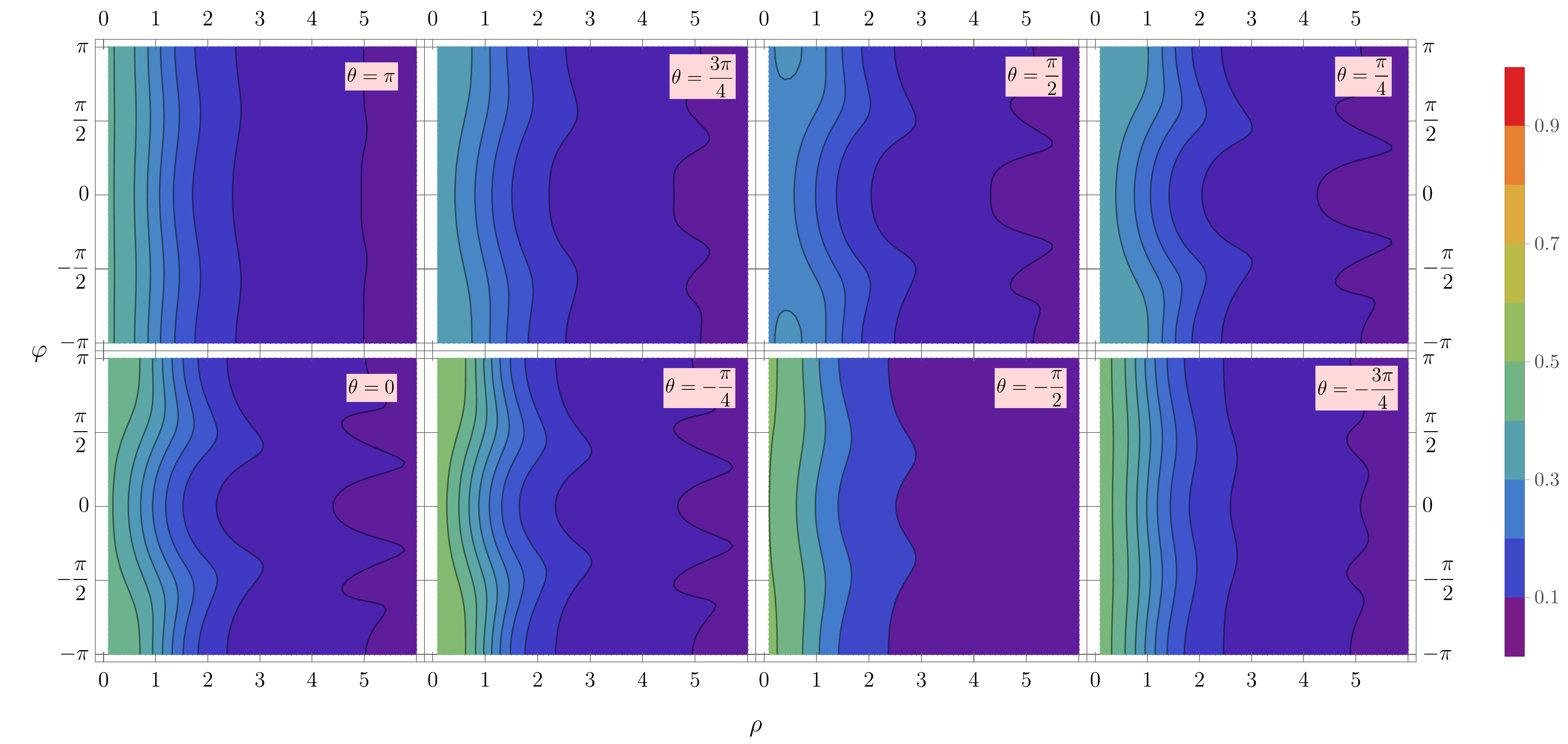}
\caption{The concurrence in the absence of the disorder, is plotted for the values of the parameters $E\equiv E/J=1$, $E/v=k=1$, $k_B\equiv B/v=0.5$.}
\label{figure_4}
\end{figure*}
\begin{figure*}[htbp]
\includegraphics*[width=0.9\linewidth]{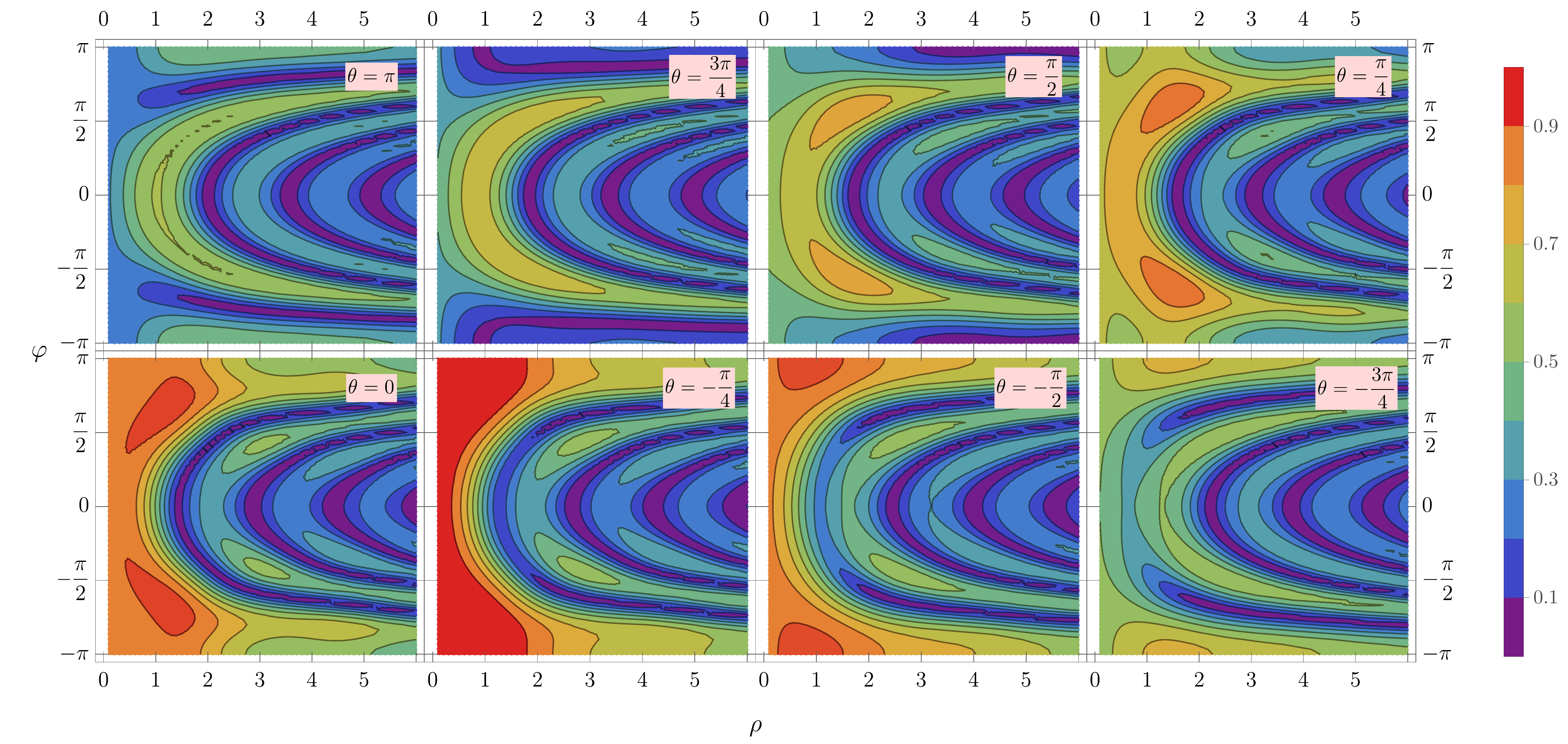}
\caption{The concurrence in the absence of the disorder, is plotted for the values of the parameters $E\equiv E/J=1$, $E/v=k=1$, $k_B\equiv B/v=1$.}
\label{figure_5}
\end{figure*}
\begin{figure*}[htbp]
\includegraphics*[width=0.9\linewidth]{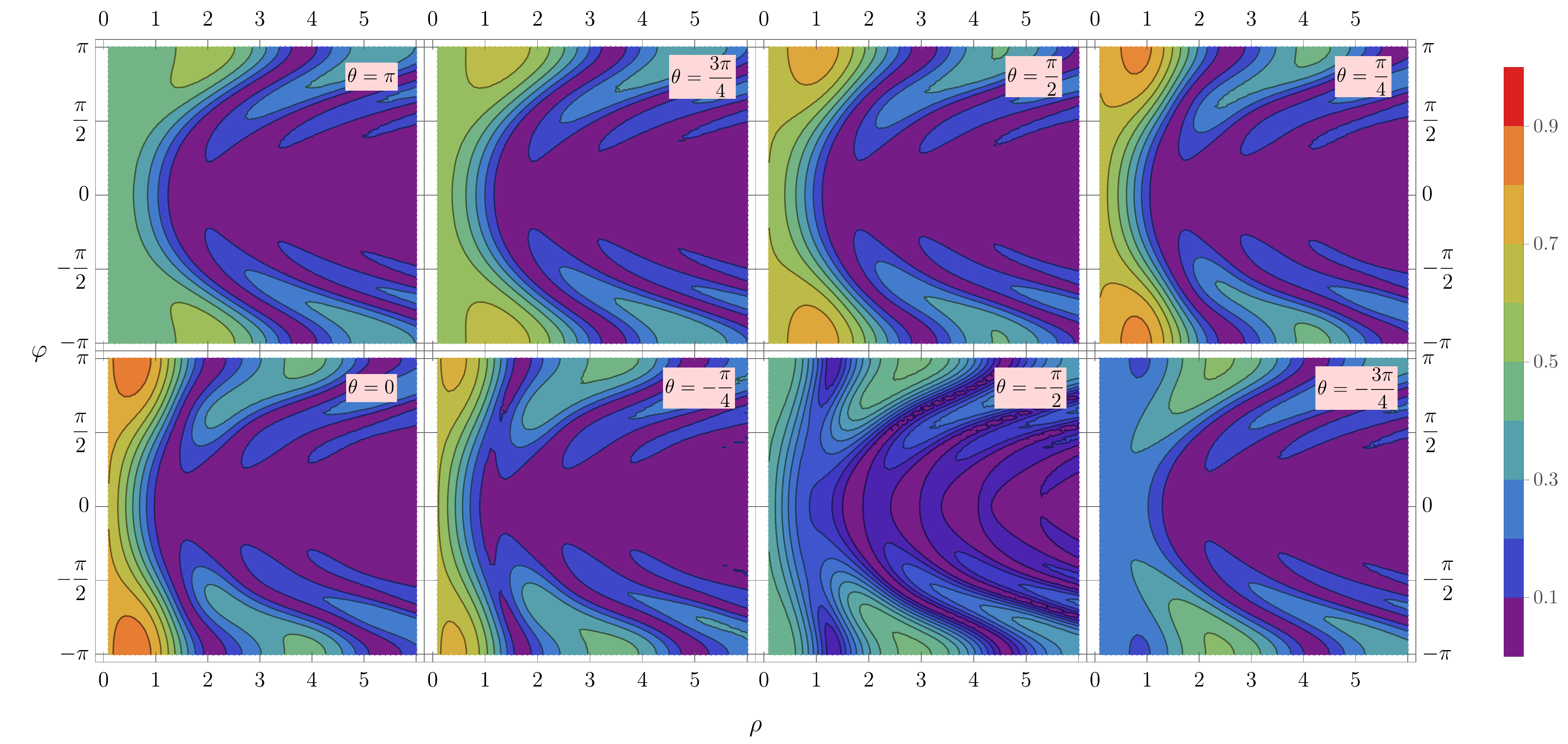}
\caption{The concurrence in the absence of the disorder, is plotted for the values of the parameters $E\equiv E/J=1$, $E/v=k=1$, $k_B\equiv B/v=1.5$.}
\label{figure_6}
\end{figure*}
An interesting fact is the dependence of entanglement on the external magnetic field. When Zeeman energy is equal to half of the energy of itinerant electron $B=E/2$, $E=J$, entanglement in the system tends to zero Fig.\ref{figure_4}.   This fact has a simple explanation if we look closely at the structure of integral equations in Eq.(28). As we see, $B=E/2$ is the particular case when the right-hand side of the second equation in Eq.(28) becomes zero disentangling the system. Entanglement increases with the increase of the magnetic field and reaches its maximum value when Zeeman splitting and the energy of itinerant electron are equal to each other $B=E$,  see Fig.\ref{figure_5}. Concurrence is larger than $\mathcal{C}>0.7$ on the distances between itinerant electron and quantum dot up to the several localization lengths $l_0$. With the further increase of the magnetic field strength $B=3/2J$, electrons disentangle again Fig.\ref{figure_6}. This fact also has a clear physical explanation. The strong magnetic field firmly fixes the spin of the electron in the quantum dot. Then spin of the localized electron is the essence of the frozen magnetic moment aligned along the external magnetic field and mimics features of a classical magnetic moment. Consequently, entanglement between the quantum object (spin of the itinerant electron) and the classical object (firmly fixed magnetic moment of the confined electron) is zero.

\section{Rényi entropy}
\label{sec:Rényi entropy}
Let us proceed to calculate Rényi entropy of the bipartite system after scattering. 
Following \cite{PhysRevA.99.052323}, we present bipartite density matrix through the Pauli strings:
\begin{eqnarray}\label{Pauli strings}
&&\hat \varrho=\frac{1}{4}\sum\limits_{\mu,\nu=0}^3\mathcal{R}_{\mu\nu}D_{\mu\nu},\nonumber\\
&&\mathcal{R}_{\mu\nu}=Tr\left[\hat\varrho D_{\mu\nu}\right],~~\hat\sigma_0=\mathcal{I}_{A,B}.\nonumber\\
&&D_{\mu\nu}=\hat\sigma^\mu_A\otimes\hat\sigma^\nu_B, 
\end{eqnarray}
where $D_{\mu\nu}$ is the Dirac matrix \cite{PhysRevA.93.062320}. After laborious calculations we deduce:
\begin{widetext}
\begin{eqnarray}
&&\mathbf{\mathcal{R}}=\nonumber\\
&&\begin{bmatrix}
1 & 2\text{Re}(C_1C_2^*+C_3C_4^*) & -2\text{Im}(C_1C_2^*+C_3C_4^*) & |C_1|^2-|C_2|^2+|C_3|^2-|C_4|^2\\
2\text{Re}(C_1C_3^*+C_2C_4^*) &  2\text{Re}(C_2C_3^*+C_1C_4^*) & 2\text{Im}(C_2C_3^*-C_1C_4^*) & 2\text{Re}(C_1C_3^*-C_2C_4^*)\\
-2\text{Im}(C_1C_3^*+C_2C_4^*) & -2\text{Im}(C_2C_3^*+C_1C_4^*) & 2\text{Re}(C_2C_3^*-C_1C_4^*) & -2\text{Im}(C_1C_3^*-C_2C_4^*)\\
|C_1|^2+|C_2|^2-|C_3|^2-|C_4|^2 & 2\text{Re}(C_1C_2^*-C_3C_4^*) & -2\text{Im}(C_1C_2^*-C_3C_4^*) & |C_1|^2-|C_2|^2-|C_3|^2+|C_4|^2
\end{bmatrix}.
\end{eqnarray}
\end{widetext}

The quantity of interest, the second Rényi entropy is given as:
\begin{eqnarray}\label{Renyi entropy}
&& S_2\left(\hat\varrho\right)=-\log_2Tr\left[ \hat \varrho^2\right].   
\end{eqnarray}
\begin{widetext}

Trace of the density matrix can be calculated through the following formula \cite{PhysRevA.99.052323}:

\begin{eqnarray}\label{Thrace of the density matrix}
&& Tr\left[ \hat \varrho^2\right]=\frac{1}{4}\bigg[ 1+3\left\langle \left(Z_u^A\right)^2\right\rangle+
3\left\langle \left(Z_u^B\right)^2\right\rangle+ 9\left\langle \left(Z_u^{AB}\right)^2\right\rangle\bigg],  
\end{eqnarray}
\end{widetext}
where 
\begin{eqnarray}\label{the following quantities}
&& Z_u^A=Tr\left[\hat U_A \hat\varrho \hat U_A^\dagger\hat\sigma_z^A\otimes \mathcal{I}_B\right],\nonumber\\
&& Z_u^B=Tr\left[\hat U_B \hat\varrho \hat U_B^\dagger\mathcal{I}_A\otimes\hat\sigma_z^B\right],\nonumber\\
&& Z_u^{AB}=Tr\left[\hat U \hat\varrho \hat U^\dagger\hat\sigma_z^A\otimes\hat\sigma_z^B\right], 
\end{eqnarray}
and ensemble average moments $\langle ...\rangle$ is done for the random SOI. Here $Z_u^{AB}$ corresponds to the process when both spins of electrons are flipped after scattering process, while $Z_u^A$ and $Z_u^B$ correspond to the processes when only one spin is flipped. 
After cumbersome calculations we deduce:
\begin{widetext}
\begin{eqnarray}\label{post-scattering density matrix}
&& \hat U \hat\varrho \hat U^\dagger\hat\sigma_z^A\otimes\hat\sigma_z^B=\frac{1}{C^2}\bigg(\vert C_1\vert^2\ket{0}\bra{0}_A\otimes\ket{0}\bra{0}_B-\vert C_2\vert^2\ket{0}\bra{0}_A\otimes\ket{1}\bra{1}_B-\vert C_3\vert^2\ket{1}\bra{1}_A\otimes\ket{0}\bra{0}_B+\nonumber\\
&& \vert C_4\vert^2\ket{1}\bra{1}_A\otimes\ket{1}\bra{1}_B-C_1C^*_2\ket{0}\bra{0}_A\otimes\ket{0}\bra{1}_B-C_1C^*_3\ket{0}\bra{1}_A\otimes\ket{0}\bra{0}_B+C_1C^*_4\ket{0}\bra{1}_A\otimes\ket{0}\bra{1}_B+\nonumber\\
&&C_2C^*_1\ket{0}\bra{0}_A\otimes\ket{1}\bra{0}_B-C_2C^*_3\ket{0}\bra{1}_A\otimes\ket{1}\bra{0}_B-C_2C^*_4\ket{0}\bra{1}_A\otimes\ket{1}\bra{1}_B+C_3C^*_1\ket{1}\bra{0}_A\otimes\ket{0}\bra{0}_B-\nonumber\\
&&C_3C^*_2\ket{1}\bra{0}_A\otimes\ket{0}\bra{1}_B-C_3C^*_4\ket{1}\bra{1}_A\otimes\ket{0}\bra{1}_B+C_4C^*_1\ket{1}\bra{0}_A\otimes\ket{1}\bra{0}_B-C_4C^*_2\ket{1}\bra{0}_A\otimes\ket{1}\bra{1}_B-\nonumber\\
&&C_4C^*_3\ket{1}\bra{1}_A\otimes\ket{1}\bra{0}_B\bigg).
\end{eqnarray}
\end{widetext}
% Here 
Taking into account Eq(\ref{post-scattering density matrix}) we deduce:
\begin{widetext}
\begin{eqnarray}\label{here two}
&& Z_u^{AB}=Tr\left[\hat U \hat\varrho \hat U^\dagger\hat\sigma_z^A\otimes\hat\sigma_z^B\right]=\frac{1}{C^2}\left(\vert C_1\vert^2-\vert C_2\vert^2-\vert C_3\vert^2+\vert C_4\vert^2\right).
\end{eqnarray}
If the spin of the quantum dot is frozen, similarly we obtain:
%\end{widetext}

%\begin{widetext}
\begin{eqnarray}\label{spin of the quantum dot is frozen}
&&\hat U_A \hat\varrho \hat U_A^\dagger\hat\sigma_z^A\otimes \mathcal{I}_B=\frac{1}{C'^2}\bigg(\vert C_1\vert^2\ket{0}\bra{0}_A\otimes\ket{0}\bra{0}_B-
\vert C_3\vert^2\ket{1}\bra{1}_A\otimes\ket{0}\bra{0}_B-\nonumber\\
&&C_1C^*_3\ket{0}\bra{1}_A\otimes\ket{0}\bra{0}_B+C_3C^*_1\ket{1}\bra{0}_A\otimes\ket{0}\bra{0}_B\bigg),\nonumber\\
&&C'=\sqrt{\vert C_1\vert^2+\vert C_3\vert^2},\nonumber\\
&&Z_u^A=Tr\left[\hat U_A \hat\varrho \hat U_A^\dagger\hat\sigma_z^A\otimes \mathcal{I}_B\right]=\frac{1}{C'^2}\bigg(\vert C_1\vert^2-\vert C_3\vert^2\bigg).
\end{eqnarray}
\end{widetext}
As a next step we need to define mean values of squares of quantities Eq.(\ref{the following quantities}). 
These mean values are defined as follows:

\begin{eqnarray} % align, split, gather
\label{we deduce the explicit expression}
&& \left\langle \left(Z_u^A\right)^2\right\rangle=\int dE\mathcal{P}(E)\left(Z_u^A(E)\right)^2,\nonumber\\
&& \left\langle \left(Z_u^B\right)^2\right\rangle=0,\nonumber\\
&& \left\langle \left(Z_u^{AB}\right)^2\right\rangle=\int dE\mathcal{P}(E)\left(Z_u^{AB}(E)\right)^2.\\\nonumber
\end{eqnarray}
Here $\mathcal{P}(v)=\frac{1}{\Delta v\sqrt{2\pi}}\exp\left[-\frac{(v-\bar v)^2}{2(\Delta v)^2}\right] $ is the distribution function of the random SO interaction constant $v$. 
\begin{figure}[htb]
\includegraphics[width=0.9\linewidth]{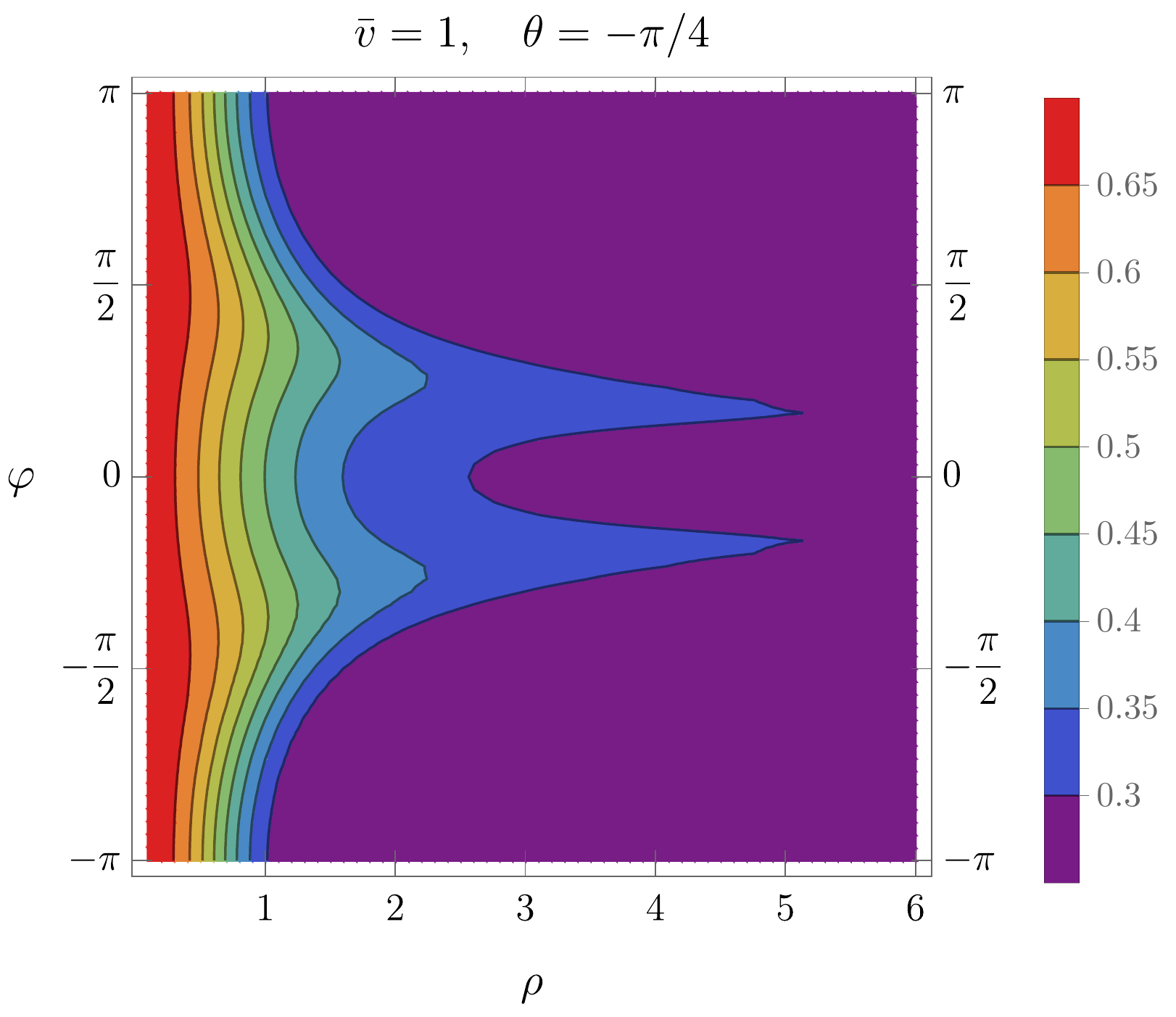}
\caption{The Rényi entropy $Tr\left[ \hat \varrho^2\right]$, in the presence of the disorder plotted for the values of the parameters $E=1$, $J=1$, $B=1$, $\theta=-\pi/4$ and the central value and the standard deviation of the Gaussian function $\mathcal{P}(v)$ accordingly $\bar v=1$ and $\Delta v=0.2$.}
\label{figure_7}
\end{figure}
The result for the Rényi entropy is plotted in Fig.\ref{figure_7}. As we see even in the presence of disorder entanglement is not zero up to the distances of several confinement length $5l_0$. Topological features of itinerant electrons sustain the formation of robust long-distance entanglement, which survives even in the presence of a strong disorder.
\begin{figure}[htb]
\includegraphics[width=0.9\linewidth]{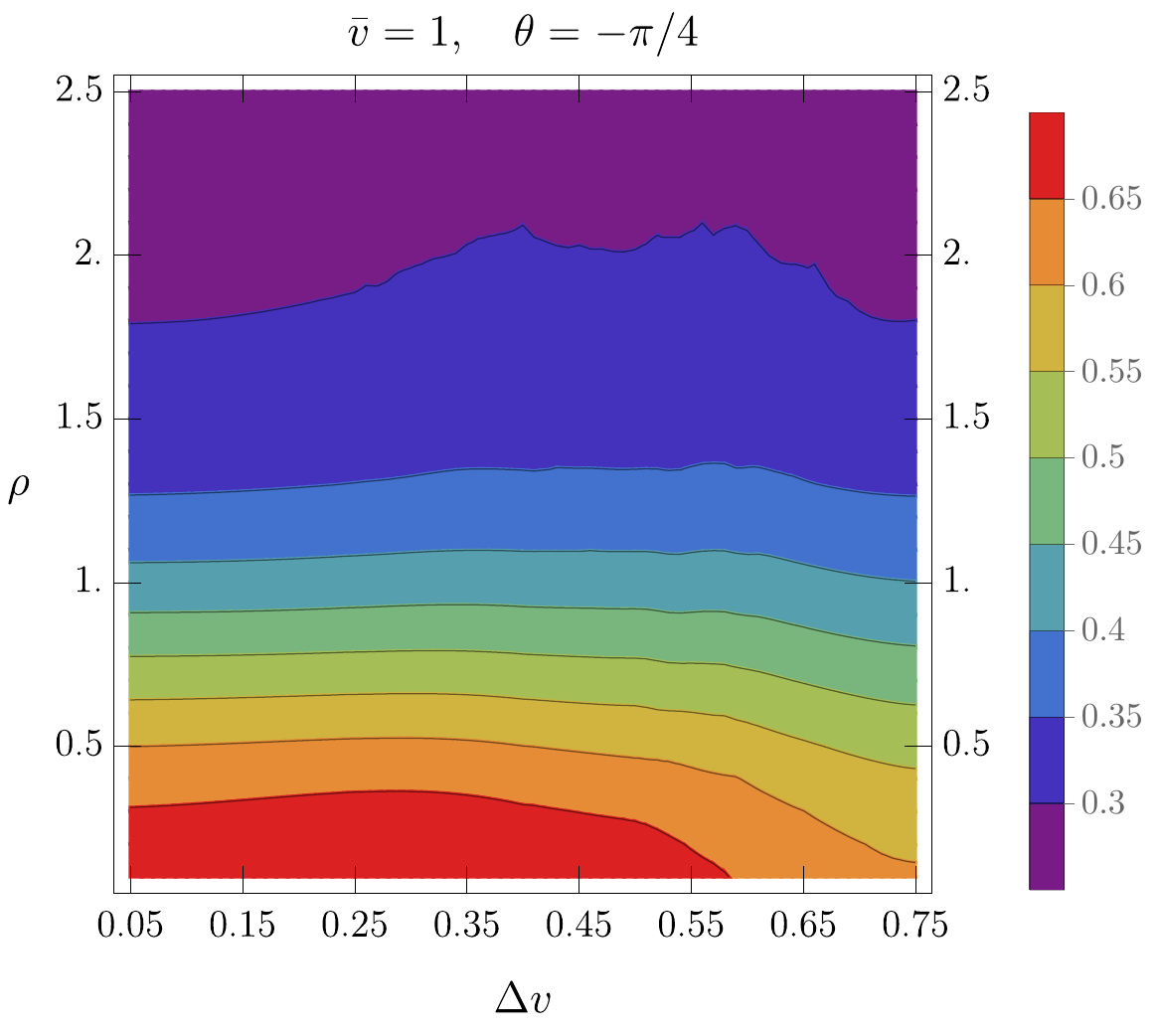}
\caption{The Rényi entropy $Tr\left[ \hat \varrho^2\right]$, as a function of the disorder strength and distance $\rho/l_0$ measured in the unit of confinement  length $l_0$. The values of parameters: $E=1$, $J=1$, $B=1$, $\theta=-\pi/4$ and the central value and the standard deviation of the Gaussian function $\mathcal{P}(v)$ accordingly $\bar v=1$.}
\label{figure_8}
\end{figure}
 Finally, in Fig. (\ref{figure_8}) we plot averaged over angle $\phi$ entropy as a function of disorder strength $\Delta v$.
 As we see from Fig. (\ref{figure_8}), entanglement decays with the strength of disorder. The decay of the entanglement increases with the strength of the disorder at small distances when the interaction between electrons is stronger. At larger distances and weaker interactions between electrons, entanglement shows robust properties concerning the strength of the disorder. However, at larger distances, the value of the entanglement is smaller.

\section{Conclusions}
\label{sec:Conclusions}
In the present work, we studied the spin-dependent scattering process in the system of two electrons when one electron localized in the quantum dot plays the role of the target for the itinerant electron of the topological insulator. Impurities in the topological insulator and alloying-induced effect of band parameter lead to the random spin-orbit interaction constant. The quantum elastic scattering process we described as a unitary process. Due to the presence of disorder and randomness
in SO constant, pre and post-scattering density matrices are connected through a unitary random spin-orbit transformation which we termed as unitary random spin-orbit Gate. We considered two witnesses of entanglement in the system, such as deterministic concurrence and ensemble-averaged Rényi entropy. We found that entanglement in the system dramatically depends on the momentum direction of the scattered electron and the external applied magnetic field. In particular, for the certain values of the magnetic field, we observed long-range entanglement when the distance between electrons substantially exceeds the confinement length $l_0$. Thus we argue that entanglement in the system can be controlled through the external magnetic field.

\section*{Acknowledgment}
This work is supported by the Grant No. FR-19-4049 from Shota Rustaveli National Science
Foundation of Georgia. It is also supported by the National Science Center in Poland as a research Project No. DEC-2017/27/B/ST3/ 02881 (SW, MI, VD). 
S.S. acknowledges support from the Norwegian Financial Mechanism under the Polish-Norwegian Research Project NCN GRIEG ‘‘2Dtronics’’, Project No. 2019/34/H/ST3/00515.

\section{Appendix}
The explicit form of the matrix elements are:
\begin{widetext}
\begin{align}\label{The potential operators}
\hat{V}_{00}(\textbf{r}_A)&=\langle\psi_{D,0}(\textbf{r}_B)|\hat{V}|\psi_{D,0}(\textbf{r}_B)\rangle=\langle\psi_{D,0}(\textbf{r}_B)|J\hat\sigma_A\hat\sigma_B\delta\left(\textbf{r}_A-\textbf{r}_B\right)|\psi_{D,0}(\textbf{r}_B)\rangle=J|\psi_D(\textbf{r}_A)|^2\hat\sigma_A^z,\nonumber\\
\hat{V}_{01}(\textbf{r}_A)&=\langle\psi_{D,0}(\textbf{r}_B)|\hat{V}|\psi_{D,1}(\textbf{r}_B)\rangle=\langle\psi_{D,0}(\textbf{r}_B)|J\hat\sigma_A\hat\sigma_B\delta\left(\textbf{r}_A-\textbf{r}_B\right)|\psi_{D,1}(\textbf{r}_B)\rangle=J|\psi_D(\textbf{r}_A)|^2(\hat\sigma_A^x-i\hat\sigma_A^y),\nonumber\\
\hat{V}_{10}(\textbf{r}_A)&=\langle\psi_{D,1}(\textbf{r}_B)|\hat{V}|\psi_{D,0}(\textbf{r}_B)\rangle=\langle\psi_{D,1}(\textbf{r}_B)|J\hat\sigma_A\hat\sigma_B\delta\left(\textbf{r}_A-\textbf{r}_B\right)|\psi_{D,0}(\textbf{r}_B)\rangle=J|\psi_D(\textbf{r}_A)|^2(\hat\sigma_A^x+i\hat\sigma_A^y),\\
\hat{V}_{11}(\textbf{r}_A)&=\langle\psi_{D,1}(\textbf{r}_B)|\hat{V}|\psi_{D,1}(\textbf{r}_B)\rangle=\langle\psi_{D,1}(\textbf{r}_B)|J\hat\sigma_A\hat\sigma_B\delta\left(\textbf{r}_A-\textbf{r}_B\right)|\psi_{D,1}(\textbf{r}_B)\rangle=-J|\psi_D(\textbf{r}_A)|^2\hat\sigma_A^z.\nonumber
\end{align}
\end{widetext}
Taking into account Eq.(\ref{The Green function}) and Eq.(\ref{The potential operators}), we rewrite  Eq.(\ref{coupled integral equations second}) in the Born approximation in the following form:
\begin{widetext}
\begin{eqnarray}
&&\begin{pmatrix} \phi_0(\textbf{r}) \\ \chi_0(\textbf{r}) \end{pmatrix} =\frac{e^{ik\rho\cos\varphi}}{\sqrt{2}}
\begin{pmatrix} 1 \\  k_+/k \end{pmatrix} -\frac{JE}{2\sqrt{2}\pi v^2}\begin{pmatrix} 1 \\  -k_+/k \end{pmatrix}\int \rho'd\rho'd\varphi'\,K_0(-iE\vert\mathbf{r}-\mathbf{r}'\vert/v)\vert\psi_D(\rho')\vert^2e^{ik\rho'\cos\varphi'}-\nonumber\\
&&\frac{JE}{2\sqrt{2}\pi v^2}\begin{pmatrix} -k_+/k \\  1 \end{pmatrix}\int \rho'd\rho'd\varphi'\,\,K_1(-iE\vert\mathbf{r}-\mathbf{r}'\vert/v)\vert\psi_D(\rho')\vert^2e^{ik\rho'\cos\varphi'},\nonumber
\end{eqnarray}
\begin{eqnarray}\label{coupled integral equations third}
&&\begin{pmatrix} \phi_1(\textbf{r}) \\ \chi_1(\textbf{r}) \end{pmatrix} =-\frac{J(E-2B)}{\sqrt{2}\pi v^2}\begin{pmatrix} k_+/k  \\  0\end{pmatrix}\int \rho'd\rho'd\varphi'\,K_0(-i(E-2B)\vert\mathbf{r}-\mathbf{r}'\vert/v)\vert\psi_D(\rho')\vert^2e^{ik\rho'\cos\varphi'}-\nonumber\\
&&\frac{J(E-2B)}{\sqrt{2}\pi v^2}\begin{pmatrix} 0 \\  k_+/k \end{pmatrix}\int \rho'd\rho'd\varphi'\,\,K_1(-i(E-2B)\vert\mathbf{r}-\mathbf{r}'\vert/v)\vert\psi_D(\rho')\vert^2e^{ik\rho'\cos\varphi'}.
\end{eqnarray}
\end{widetext}
Here ${\bf r}=(\rho\cos\varphi,\rho\sin\varphi)$, $|\mathbf{r}-\mathbf{r}'|=\sqrt{\rho^2+\rho'^2-2\rho\rho'\cos(\varphi-\varphi')}$, the wave function of the electron in the quantum dot $\psi_D(\rho')=\frac{1}{l_B\sqrt{\pi}}\exp\left[-\frac{\rho'^2}{2l_B^2}\right]$, and for the sake of simplicity we adopt the asymptotic form
$K_{l=0,1}(z)=\frac{\pi}{2z}e^{-z}\left\lbrace1+\frac{4 l^2-1}{8z}+\mathcal{O}(1/z^2)\right\rbrace $. Taking into account Eq.~(\ref{coupled integral equations third}), two electron function 
after scattering, we express in terms of coefficients:
\begin{widetext}
\begin{eqnarray}\label{notations for the coefficients}
&& A_{0}(\mathbf{r})=\int \rho'd\rho'd\varphi'\,K_0(-iE\vert\mathbf{r}-\mathbf{r}'\vert/v)\vert\psi_D(\rho')\vert^2e^{ik\rho'\cos\varphi'},\nonumber\\
&& A_{1}(\mathbf{r})=\int \rho'd\rho'd\varphi'\,\,K_1(-iE\vert\mathbf{r}-\mathbf{r}'\vert/v)\vert\psi_D(\rho')\vert^2e^{ik\rho'\cos\varphi'},\nonumber\\
&&A_{0B}(\mathbf{r})=\int \rho'd\rho'd\varphi'\,K_0(-i(E-2B)\vert\mathbf{r}-\mathbf{r}'\vert/v)\vert\psi_D(\rho')\vert^2e^{ik\rho'\cos\varphi'},\nonumber\\
&&A_{1B}(\mathbf{r})=\int \rho'd\rho'd\varphi'\,\,K_1(-i(E-2B)\vert\mathbf{r}-\mathbf{r}'\vert/v)\vert\psi_D(\rho')\vert^2e^{ik\rho'\cos\varphi'}.
\end{eqnarray}
\end{widetext}

In the asymptotic case $kr\gg 1$ when distance between particles exceeds the magnetic localization length $r>l_B$ we can analytically perform integration over angle in Eq.(\ref{notations for the coefficients}) is:
\begin{widetext}
\begin{eqnarray}\label{integration over angle}
&& A_{l}(\mathbf{r})=-\frac{i v}{2El^2_B}\int\limits_0^\infty\frac{\rho' d\rho'}{\sqrt{\rho^2+\rho'^2}}\frac{\exp\left[-\frac{iE}{v}\sqrt{\rho^2+\rho'^2-\rho\rho'(z_1+1/z_1)}\right]}{\sqrt{2}(1-a^2)^{1/4}}\exp\left[-\frac{\rho'^2}{l^2_B}\right]\times\nonumber\\
&&\exp\left[\frac{ik\rho'}{2}\left((z_1+1/z_1)\cos\varphi+i(z_1-1/z_1)\sin\varphi\right)\right]-\nonumber\\
&&\frac{(4 l^2-1)v^2}{16E^2l^2_B}\int\limits_0^\infty\frac{\rho' d\rho'}{\rho^2+\rho'^2}\frac{\exp\left[-\frac{iE}{v}\sqrt{\rho^2+\rho'^2-\rho\rho'(z_1+1/z_1)}\right]}{\sqrt{2}(1-a^2)^{1/4}}\exp\left[-\frac{\rho'^2}{l^2_B}\right]\times\nonumber\\
&&\exp\left[\frac{ik\rho'}{2}\left((z_1+1/z_1)\cos\varphi+i(z_1-1/z_1)\sin\varphi\right)\right],\nonumber\\
&&A_{l B}(\mathbf{r})=\frac{i v}{2(E-2B)l^2_B}\int\limits_0^\infty\frac{\rho' d\rho'}{\sqrt{\rho^2+\rho'^2}}\frac{\exp\left[-\frac{i(E-2B)}{v}\sqrt{\rho^2+\rho'^2-\rho\rho'(z_1+1/z_1)}\right]}{\sqrt{2}(1-a^2)^{1/4}}\exp\left[-\frac{\rho'^2}{l^2_B}\right]\times\nonumber\\
&&\exp\left[\frac{ik\rho'}{2}\left((z_1+1/z_1)\cos\varphi+i(z_1-1/z_1)\sin\varphi\right)\right]-\nonumber\\
&&\frac{(4 l^2-1)v^2}{16(E-2B)^2l^2_B}\int\limits_0^\infty\frac{\rho' d\rho'}{\rho^2+\rho'^2}\frac{\exp\left[-\frac{i(E-2B)}{v}\sqrt{\rho^2+\rho'^2-\rho\rho'(z_1+1/z_1)}\right]}{\sqrt{2}(1-a^2)^{1/4}}\exp\left[-\frac{\rho'^2}{l^2_B}\right]\times\nonumber\\
&&\exp\left[\frac{ik\rho'}{2}\left((z_1+1/z_1)\cos\varphi+i(z_1-1/z_1)\sin\varphi\right)\right].\nonumber\\
\end{eqnarray}
\end{widetext}
Here we introduced the notations: $z_1=\frac{1}{a}+\sqrt{\frac{1}{a^2}-1}$,
$z_2=\frac{1}{a}-\sqrt{\frac{1}{a^2}-1}$, and $a=\frac{2\rho\rho'}{\rho^2+\rho'^2}$.

\bibliography{TEXT_KK_VD.LC}

\end{document}